\documentclass[12pt,preprint]{aastex}
\usepackage{psfig}

\newcommand\pp     {$\pm$}

\def\degr{\hbox{$^\circ$}}
\newcommand\Lunit   {erg s$^{-1}$}
\newcommand\funit   {erg cm$^{-2}$ s$^{-1}$}
\def\degr{\hbox{$^\circ$}}

\def\la{\mathrel{\mathchoice {\vcenter{\offinterlineskip\halign{\hfil
$\displaystyle##$\hfil\cr<\cr\noalign{\vskip1.5pt}\sim\cr}}}
{\vcenter{\offinterlineskip\halign{\hfil$\textstyle##$\hfil\cr<\cr
\noalign{\vskip1.0pt}\sim\cr}}}
{\vcenter{\offinterlineskip\halign{\hfil$\scriptstyle##$\hfil\cr<\cr
\noalign{\vskip0.5pt}\sim\cr}}}
{\vcenter{\offinterlineskip\halign{\hfil$\scriptscriptstyle##$\hfil
\cr<\cr\noalign{\vskip0.5pt}\sim\cr}}}}}

\righthead{GRO J1744--28 in quiescence}
\slugcomment{Re-submitted to ApJ Letters: February 12, 2002}

\begin{document}

\title{A {\itshape Chandra} observation of GRO J1744--28: the bursting
pulsar in quiescence}

\author{Rudy Wijnands\altaffilmark{1,2}, Q. Daniel Wang\altaffilmark{3}}

\altaffiltext{1}{Center for Space Research, Massachusetts Institute of
Technology, 77 Massachusetts Avenue, Cambridge, MA 02139-4307, USA;
rudy@space.mit.edu}

\altaffiltext{2}{Chandra Fellow}

\altaffiltext{3}{Astronomy Department, University of Massachusetts,
Amherst, MA 01003, USA}

\begin{abstract}
We present a {\it Chandra}/ACIS-I observation of GRO J1744--28. We
detected a source at a position of R.A = 17$^h$ 44$^m$ 33.09$^s$ and
Dec. = --28\degr~44$'$ 27.0$''$ (J2000.0; with a 1$\sigma$ error of
$\sim$0.8 arcseconds), consistent with both {\it ROSAT} and
interplanetary network localizations of GRO J1744--28 when it was in
outburst.  This makes it likely that we have detected the quiescent
X-ray counterpart of GRO J1744--28. Our {\it Chandra} position
demonstrates that the previously proposed infrared counterpart is not
related to GRO J1744--28. The 0.5--10 keV luminosity of the source is
$2 - 4 \times10^{33}$ \Lunit~(assuming the source is near the Galactic
center at a distance of 8 kpc). We discuss our results in the context
of the quiescent X-ray emission of pulsating and non-pulsating neutron
star X-ray transients.

\end{abstract}


\section{Introduction \label{section:intro}}

X-ray transients sporadically exhibit very bright outbursts during
which their X-ray luminosity can be as high as $10^{36-39}$
\Lunit. However, most of their time they are in their quiescent state
during which they emit X-rays only at a level of $10^{30-34}$ \Lunit.
The mechanisms behind this quiescent X-ray emission are still not
understood (see, e.g., Menou et al. 1999; Campana \& Stella 2000;
Bildsten \& Rutledge 2002).  The most dominant model for the quiescent
properties of non-pulsating neutron star transients, which contain a
neutron star with a very low magnetic field ($<10^{10}$ Gauss), is the
one which assumes that the X-rays below a few keV are due to the
cooling of the neutron star after the accretion has stopped (see,
e.g., Brown, Bildsten, \& Rutledge 1998 and references therein).  In
the quiescent X-ray spectra of several of those systems an extra power
law component above a few keV is present (see, e.g., Asai et al. 1996,
1998; Campana et al. 1998a; Rutledge et al. 2001), but the nature of
this component is even more unclear (see, e.g., Campana \& Stella 2000
for a discussion).

Detailed studies of the quiescent emission from transient X-ray
pulsars (with a neutron star magnetic field strength of $>10^{11}$
Gauss) have been inhibited by the lack of detected systems. So far,
only two X-ray pulsars have been detected in quiescence, A 0535+26
(Negueruela et al. 2000) and 4U 0155+63 (Campana et al. 2001). Their
X-ray luminosities were consistent with the predictions of the cooling
neutron star model for the non-pulsating systems, suggesting that also
this model might apply to the pulsating ones. However, due to the low
statistics of the data a detailed testing of the model could not be
performed. To get more insight in the quiescent emission of X-ray
pulsars those systems have to be observed with higher sensitivity
instruments like {\it Chandra} or {\it XMM-Newton}. Furthermore, more
quiescent X-ray pulsars have to be detected to determine if all of
those systems are consistent with the cooling model or that some
systems might have different properties which would suggest
alternative X-ray production mechanisms (e.g., accretion down to the
magnetospheric radius; Stella et al. 1994; Corbet 1996). A good
candidate for such studies is the ``bursting pulsar'' GRO J1744--28.

GRO J1744--28 was discovered in December 1995 with the Burst and
Transient Source Experiment (BATSE) aboard the {\it Compton Gamma-Ray
Observatory} ({\it CGRO}; Fishman et al. 1995; Kouveliotou et
al. 1996). The source exhibited rapidly repeating, very bright X-ray
bursts which are likely due to accretion disk instabilities (e.g.,
Lewin et al. 1996). The source also exhibited pulsed emission with a
pulsation frequency of 2.1 Hz (Finger et al. 1996). Its bursting and
pulsating nature lead to the source being called the bursting
pulsar. So far, two major outbursts have been detected, one starting
in December 1995 and the other one a year later in December 1996. The
latter one ended in April 1997 (see, e.g., Woods et al. 1999).

Augusteijn et al. (1997) likely detected GRO J1744--28 using a {\it
ROSAT} observation performed in March 1996. This observation showed a
bright source with a luminosity of $\sim 2 \times 10^{37}$ \Lunit~(for
an assumed distance\footnote{The distance to the source is unknown but
the high column density measured by Nishiuchi et al. (1999) and the
proximity of the source on the sky to the center of the Galaxy make a
large distance likely.} of 8 kpc; 0.1--2.4 keV).  Archival {\it ROSAT}
observations did not detect a source on this position with an upper
limit on the luminosity of a few times $10^{33}$ \Lunit~(0.1--2.4 keV;
Augusteijn et al. 1997). This transient nature of the source makes it
very likely that indeed the {\it ROSAT} source is GRO J1744--28
despite that no bursts or pulsations were detected (note that the
large {\it ROSAT} upper limit on the pulsations obtained by Augusteijn
et al. 1997 was completely consistent with the strength of the
pulsations as measured with BATSE and the {\it Rossi X-ray Timing
Explorer} [{\it RXTE}]). Localization of GRO J1744--28 by
triangulating the data obtained for this source with {\it Ulysses} and
BATSE (part of the interplanetary network [IPN]) resulted in a
position (Hurley et al. 2000) which partly overlapped that of the {\it
ROSAT} error circle, confirming the identification of the {\it ROSAT}
source with GRO J1744--28. Cole et al. (1997) and Augusteijn et
al. (1997) identified a possible infrared counterpart at the edge of
the {\it ROSAT} error circle, although its enigmatic properties
spurred the suggestion that it might be an instrumental artifact
(Augusteijn et al. 1997; but see Cole et al. 1997).

Here we present a {\it Chandra}/ACIS-I observation of the region
containing GRO J1744--28. We discovered a weak source near the center
of the {\it ROSAT} error circle, which is likely the quiescent
counterpart of the {\it ROSAT} transient and therefore likely of GRO
J1744--28.

\section{Observation, analysis, and results}

GRO J1744--28 was in the field of view ($\sim$7.1 arcminutes off-axis)
during one of the observations which were obtained as part of the {\it
Chandra} survey of the Galactic Center region (Wang, Gotthelf, \& Lang
2002). This particular observation was performed on 2001 July 18 17:37
- 20:49 UT with an exposure time of $\sim10.6$ ksec. The ACIS-I
instrument was used during this observation. To limit the telemetry
rate only those photons with energy above 1 keV were transmitted to
Earth. The data were analysed using the analysing packet CIAO, version
2.2.1, and the threats listed on the CIAO web pages\footnote{Available
at http://asc.harvard.edu/ciao/}.

\begin{figure}
\begin{center}
\begin{tabular}{c}
\psfig{figure=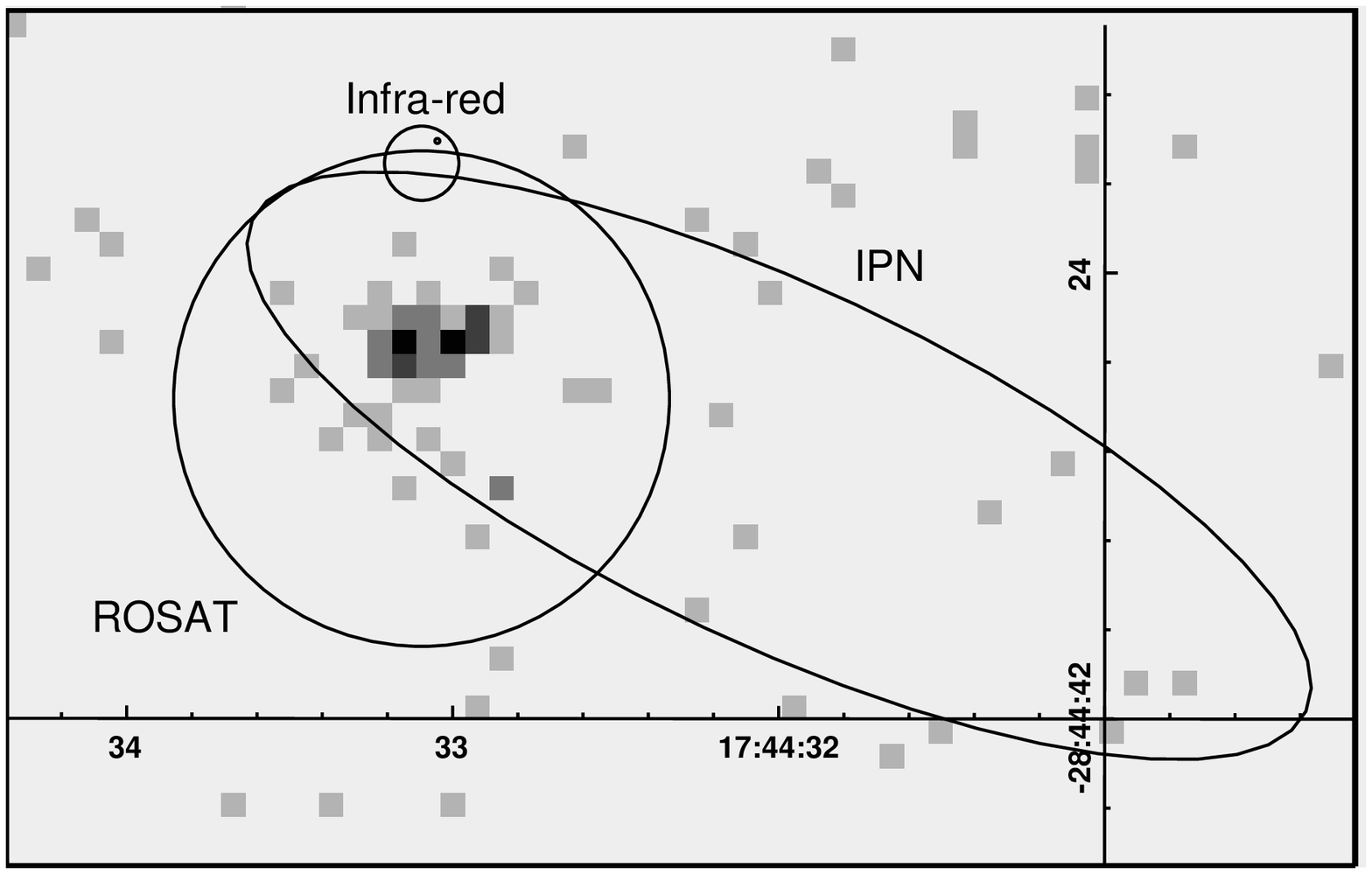,width=16cm}
\end{tabular}
\figcaption{The {\it Chandra}/ACIS-I image (for photons with energy of
1 keV or higher) of GRO J1744--28, rebinned by a factor of 2. Due to
the point-spread-function for off-axis point sources, the source
appears to be extended, but it is consistent with being a point
source. The coordinates are for epoch J2000.0.  Also shown are the
{\it ROSAT} error circle (Augusteijn et al. 1997), the IPN
localization ellipse (Hurley et al. 2000) and the error circles of the
possible infrared counterpart (Augusteijn et al. 1997: large; Cole et
al. 1997: small). It is obvious that the {\it Chandra} source is
consistent with the {\it ROSAT} and the IPN source, but not with the
infrared source.\label{fig:image} }
\end{center}
\end{figure}

The resulting image of the region near GRO J1744--28 is shown in
Figure~\ref{fig:image}. We detected one source in the {\it ROSAT}
error circle (Augusteijn et al. 1997) and the IPN error ellipse
(Hurley et al. 2000) of GRO J1744--28.  This strongly suggests that
indeed we have detected the quiescent counterpart of GRO J1744--28.
The best source position (as obtained with the CIAO tool WAVDETECT) is
R.A. = 17$^h$ 44$^m$ 33.09$^s$ and Dec. = --28\degr~44$'$ 27.0$''$
(for J2000.0). Due to the low number of detected photons the
statistical error on the source position is 0.3 arcseconds. However,
the satellite pointing error is approximately 0.7 arcseconds
(1$\sigma$; Aldcroft et al. 2000) and dominates the positional
inaccuracy. The proposed infrared counterpart (Cole et al. 1997;
Augusteijn et al. 1997) is not consistent with our {\it Chandra}
position.

The elongated structure of the detected source in
Figure~\ref{fig:image} is due to the extended point-spread-function
for a point source approximately 7 arcminutes off-axis. We detected
52\pp8 counts (corrected for background) from the source position,
resulting in a count rate of 4.9\pp0.7 $\times10^{-3}$ counts
s$^{-1}$. The source spectrum was extracted using a circle with a
radius of 10$''$ on the source position\footnote{For a $\sim7'$ off
axis source an extraction radius of 10$''$ ensures that virtually all
source photons are extracted and that no contamination of source
photons occurs in the background region.}. The background data were
obtained by using an annulus on the same position with an inner radius
of 10$''$ and an outer one of 30$''$. The data were rebinned using the
FTOOLS routine grppha into bins with a minimum of 5 counts per bin. We
employed both the $\chi^2$ and CASH (Cash 1979) statistics to fit the
data. Both methods give very similar results and we will only discuss
the results obtained with the $\chi^2$ method.  The obtained spectrum
was fitted using XSPEC version 11.1.0 (Arnaud 1996).

\begin{figure}
\begin{center}
\begin{tabular}{c}
\psfig{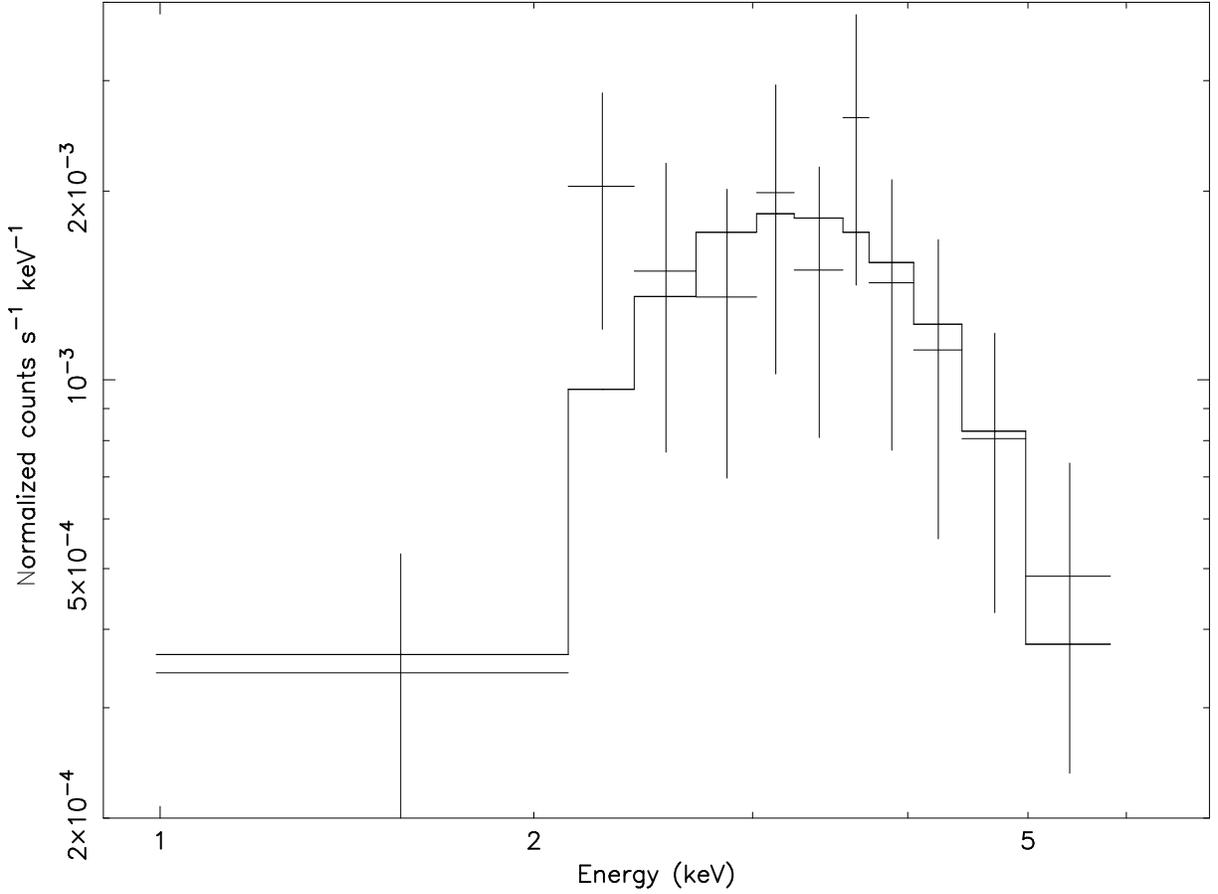}
\end{tabular}
\figcaption{The energy spectrum (for photons above 1 keV) of GRO
J1744--28 as measured with the {\it Chandra}/ACIS-I instrument. The
solid line through the data is the best blackbody fit.
\label{fig:spectra} }
\end{center}
\end{figure}

The spectrum (Fig.~\ref{fig:spectra}) was of poor quality and every
single-component model provided an acceptable fit ($\chi^2/dof \la
1$).  The column density could not be constrained and was fixed to the
value as determined using {\it ASCA} data ($N_{\rm H} \sim 5.5 \times
10^{22}$ cm$^{-2}$; Nishiuchi et al. 1999) during times when the
source was in outburst.  A pure power-law model resulted in a photon
index of 2.4\pp0.7 (all fit parameters are for a 95\% confidence
level) and a flux of $5.3\times10^{-13}$ \funit~(0.5--10 keV;
unabsorbed; all subsequent fluxes are for this energy range and
unabsorbed). A blackbody model resulted in a temperature $kT$ of
0.8\pp0.2 keV, with a radius of the emitting region of only
0.2$^{+1.6}_{-0.9}$ km (assuming a distance of 8 kpc), and a flux of
$2.5\times10^{-13}$ \funit. When fitting a neutron star hydrogen
atmosphere model (Zavlin, Pavlov, \& Shibanov 1996) to the data, the
radius of the emitting region could not be constrained and was fixed
to 10 km. The obtained temperature at infinity $kT_\infty$ was
0.4\pp0.2 keV and the resulting flux is $3.0\times10^{-13}$
\funit. The uncertainties in which spectral model to use resulted in a
possible range for the flux of $2.5 - 5.3 \times 10^{-13}$ \funit.  If
the source is indeed near the Galactic center at a distance of $\sim8$
kpc, then the 0.5--10 keV luminosity would be in the range of $2 - 4
\times 10^{33}$ \Lunit.

When the quiescent spectrum of non-pulsating neutron star X-ray
transients are fit with a single component model consisting of either
a blackbody or a neutron star atmosphere model, the obtained
temperatures are usually 0.2--0.3 keV (using a blackbody model; e.g.,
Bildsten \& Rutledge 2002 and references therein) or $\sim0.1$ keV
(using a neutron star atmosphere model). Our spectrum of GRO J1744--28
is not consistent with such low temperatures ($\chi^2/dof > 3$): both
the blackbody and the neutron star atmosphere model fall-off more
rapidly at energies above 1 keV than our data.  However, several of
those non-pulsating systems have displayed a composite quiescent X-ray
spectrum, with a soft, most likely thermal component below $\sim1$ keV
and a hard, power-law type component at higher energies (e.g., Asai et
al. 1996, 1998). We have fitted our data using such a composite
spectrum consisting of either a blackbody or a neutron star atmosphere
model for the soft component and a power-law with photon index of 1 or
2 for the hard component. However, due to the low number of photons
detected and the high column density towards the source, no
constraints could be set on the temperature of the soft component. The
data are fully consistent with the temperatures observed for other
systems (e.g., a blackbody temperature of 0.2--0.3 keV) and with an
upper limit on the thermal flux of $3 \times 10^{-13}$ \funit~(0.5--10
keV; for a $kT$ of 0.3 keV) or $9 \times 10^{-13}$ \funit~(for a $kT$
of 0.2 keV).

The detected source was the only one present in the {\it ROSAT} error
circle and the IPN ellipse. Hands et al. (2002) reported on {\it
XMM-Newton} observations performed on the Galactic plane to study the
Galactic X-ray point source population. Using their log $N$ -- log $S$
curve and a 2--6 keV flux of $\sim 1 - 2\times10^{-13}$ \funit~for GRO
J1744--28, we estimate that about 10 to 20 sources per square degree
can be detected with the same flux. This gives a probability of $\sim1
- 2\times10^{-4}$ that a random field source would fall in the surface
area traced out by the intersection of the {\it ROSAT} error circle
and the IPN error ellipse. The source density will likely be higher in
the Galactic center region which would increase this probability. But
the calculated probability can be used as a first approximation and it
indicates that the detected source is likely GRO J1744--28. However,
if the detected {\it Chandra} source is not GRO J1744--28, then the
flux upper limit on this source would be about $2 - 5 \times 10^{-14}$
\funit~(depending on which spectral model is assumed).  Our
observation had a time resolution of only $\sim3.2$ seconds, which did
not allow us to search for the 2.1 Hz pulsations. The low number of
photons detected did not allow for a stringent conclusion on the
possible variability of the source on longer time scales.

\section{Discussion\label{section:discussion}}

We detected a {\it Chandra} source in the {\it ROSAT} error circle
(Augusteijn et al. 1997) and IPN ellipse (Hurley et al. 2000) of GRO
J1744--28, making it likely that we have detected the quiescent X-ray
counterpart of GRO J1744--28. The proposed infrared counterpart (Cole
et al. 1997; Augusteijn et al. 1997) is not consistent with our {\it
Chandra} position and the nature of this source is unclear (it might
be, as suggested by Augusteijn et al. 1997, an artifact, although Cole
et al. 1997 could not confirm this).  Our quiescent flux of $2.5 - 5.3
\times 10^{-13}$ \funit~(0.5--10 keV; unabsorbed) is consistent with
the non-detection of GRO J1744--28 during the archival {\it ROSAT}
observation reported by Augusteijn et al. (1997; using their count
rate upper limit for GRO J1744--28, an 0.5--10 keV unabsorbed flux
upper limit range can be derived using PIMMS\footnote{Available at
http://heasarc.gsfc.nasa.gov/Tools/w3pimms.html} of
$7-10\times10^{-13}$ \funit, depending on the spectral model).

One possible mechanism producing the quiescent emission is residual
accretion on to the surface of the neutron star. However, Cui (1997)
argued (based on the sudden decrease of the pulsation amplitude at
times when the 2--60 keV flux of GRO J1744--28, as measured with {\it
RXTE}, was below $\sim2\times 10^{-9}$ \funit) that GRO J1744--28 is
in the ``propeller'' regime at relatively low fluxes. In this regime,
the magnetic field of the neutron star inhibits accretion on to the
neutron star surface. Our {\it Chandra} flux for GRO J1744--28 is much
lower than the critical flux limit given by Cui (1997) strongly
suggesting that the source was in the propeller regime during our
observation.  Therefore, accretion on to the surface of the neutron
star is not likely to be the cause behind the quiescent X-ray emission
of GRO J1744--28. Similar conclusions were also reached for the
quiescent emission from the transient X-ray pulsars A 0535+26
(Negueruela et al. 2000) and 4U 0155+63 (Campana et al. 2001).

A possible mechanism to produce the observed quiescent X-rays when the
source is in the propeller regime might be accretion down to the
magnetospheric radius $r_{\rm m}$ (e.g., Stella et al. 1994; Corbet
1996; Campana et al. 1998b), which is approximately give by $(GM_{\rm
ns})^{-1/7} \mu^{4/7} \dot{M}^{-2/7}$, in which $M_{\rm ns}$ is the
neutron star mass, $\mu$ the neutron star magnetic moment, and
$\dot{M}$ the accretion rate. The luminosity produced will be $L = G
M_{\rm ns} \dot{M}/r_{\rm m}$. Using our measured luminosity and the
magnetic field strength found by Cui (1997; $\sim2\times10^{11}$
Gauss), then $\dot{M} \sim 3 - 5 \times 10^{15}$ g s$^{-1}$ and
$r_{\rm m} \sim 2 - 3 \times 10^{8}$ cm.  This radius is larger than
the corotation radius $r_{\rm c} = (GM_{\rm ns})^{1/3}(P_{\rm
spin}/2\pi)^{2/3}$ (with $P_{\rm spin}$ the neutron star spin period),
which is $\sim10^{8}$ cm for GRO J1744--28, and therefore it fulfills
the condition that the source should be in the propeller
regime. Although a rough estimate\footnote{Large uncertainties are
present in the exact luminosity of the source (due to uncertainties in
the spectral shape and in the distance to the source), its magnetic
field strength, and in certain unknown constants (like the accretion
efficiency), which have been assumed to be roughly 1. A full detailed
discussion of those uncertainties is beyond the scope of this {\it
Letter}.}, it indicates that accretion down to the magnetospheric
radius might be able to produce the quiescent X-ray luminosity of GRO
J1744--28. However, contributions to the X-ray luminosity might also
come from several other mechanisms, like accretion onto the neutron
star surface (possibly on localized areas such as the magnetic poles)
due to the leakage of matter through the magnetospheric barrier.

Another contribution to the quiescent X-ray emission will be the
thermal X-ray emission from the neutron star surface which should give
a rock bottom lower limit on the X-ray luminosity. Due to the low
statistics of our data and the high column density towards GRO
J1744--28, we are not able to accurately probe such a thermal
component. However, if the temperature of the thermal component is
similar to what has been observed for the quiescent spectra of the
non-pulsating neutron star transients, then at least the detected
emission in our {\it Chandra} observation is mostly due to an another
component, which has its origin probably in one of the above discussed
mechanisms.

Progress in our understanding of the quiescent X-ray emission can be
made for GRO J1744--28 by obtaining longer {\it Chandra} or {\it
XMM-Newton} observations of this source. Such observations will be
able to constrain its luminosity and spectral shape much better than
we are able to due with the present {\it Chandra} data and will
determine if the quiescent spectrum of GRO J1744--28 is similar to
that of the quiescent non-pulsating systems, or that a fundamental
difference is present. Detecting the pulsations in observations with
sufficient time resolution will also constrain the mechanisms for the
quiescent X-rays in this system.  Detecting the pulsations in
quiescence would also unambiguously proof that the detected source is
GRO J1744--28.

{\it Note added in manuscript:} After submission of our paper, we
became aware of the paper by Daigne et al. (2002) who reported on a
{\it XMM-Newton} observation of GRO J1744--28 in quiescence. Their
reported position and X-ray flux for the source are consistent with
ours.

\acknowledgments

RW was supported by NASA through Chandra Postdoctoral Fellowship grant
number PF9-10010 awarded by CXC, which is operated by SAO for NASA
under contract NAS8-39073. WQD is supported by the CXC grant
SAO-GO1-2150A. We thank Jon Miller for comments on a previous version
of this letter.

\end{document}